\documentclass{ceurart}

\usepackage{cleveref}
\usepackage{tablefootnote}
\usepackage{glossaries}
\usepackage{hhline}
\usepackage{todonotes}

\usepackage{amsmath}
\DeclareMathOperator*{\argmax}{arg\,max}

\newacronym{un}{UN}{United Nations}
\newacronym{eu}{EU}{European Union}
\newacronym{sdg}{SDG}{Sustainable Development Goal}
\newacronym{abm}{ABM}{Agent-Based Model}
\newacronym{ig}{IG}{Institutional Grammar}

\begin{document}

\copyrightyear{2021}
\copyrightclause{Copyright for this paper by its authors.
  Use permitted under Creative Commons License Attribution 4.0
  International (CC BY 4.0).}

\conference{AMPM 2022: 2nd Workshop in Agent-based Modeling \& Policy-Making, JURIX 2022, December 14, Saarbr\"{u}cken, Germany}

\title{An Agent-Based Model for Poverty and Discrimination Policy-Making}

\author[1,0]{Nieves Montes}[orcid=0000-0001-6981-7893,email=nmontes@iiia.csic.es]
\author[2,0]{Georgina Curto}[orcid=0000-0002-8232-3131,email=gcurtore@nd.edu]
\author[1]{Nardine Osman}[orcid=0000-0002-2766-3475,email=nardine@iiia.csic.es]
\author[1]{Carles Sierra}[orcid=0000-0003-0839-6233,email=sierra@iiia.csic.es]

\address[1]{Artificial Intelligence Research Institute, IIIA-CSIC, Barcelona, Spain}
\address[2]{University of Notre Dame, Notre Dame, USA}
\address[0]{Both authors have contributed equally to this paper.}

\begin{abstract}
    The deceleration of global poverty reduction in the last decades suggests that traditional redistribution policies are losing their effectiveness. Alternative ways to work towards the \#1 United Nations Sustainable Development Goal (poverty eradication) are required. NGOs have insistingly denounced the criminalization of poverty, and the social science literature suggests that discrimination against the poor (a phenomenon known as \emph{aporophobia}) could constitute a brake to the fight against poverty. This paper describes a proposal for an agent-based model to examine the impact that aporophobia at the institutional level has on poverty levels. This aporophobia agent-based model (AABM) will first be applied to a case study in the city of Barcelona. The regulatory environment is central to the model, since aporophobia has been identified in the legal framework. The AABM presented in this paper constitutes a cornerstone to obtain empirical evidence, in a non-invasive way, on the causal relationship between aporophobia and poverty levels. The simulations that will be generated based on the AABM have the potential to inform a new generation of poverty reduction policies, which act not only on the redistribution of wealth but also on the discrimination of the poor. 
\end{abstract}


\maketitle

\section{Introduction}\label{sec:intro}

Global poverty reduction has been slowing down in the last decades \cite{Claudia2018}, and the impact of the Covid-19 pandemic could make poverty levels escalate sharply by up to 8.3\% \cite{UnitedNations2022}. This setback is currently aggravated by rising inflation and the war in Ukraine. The \acrfull{un} informs that currently 676M people live in extreme poverty, and an additional 75M to 95M people could have fallen into this category in 2022. Only within the 27 countries of the \acrlong{eu} (\acrshort{eu}-27), 95.4M people are currently at risk of poverty or social exclusion, about 20\% of the population \cite{Eurostat2022}. In this context, there is a growing consensus that using traditional wealth redistribution policies as the sole weapon to fight poverty is insufficient. Alternative and innovative solutions are needed to achieve the \#1 \acrshort{un} \acrfull{sdg}: poverty eradication.

Strikingly, the phenomenon of discrimination against the poor has not received the deserved attention in the literature. In 1995 philosopher Adela Cortina coined the term \emph{aporophobia} to refer to it \cite{Cortina1995}. In 2021, the Spanish legal framework pioneered the inclusion of this form of discrimination as an aggravating factor for hate crimes \cite{BoletinOficialdelEstadoMinisteriodelaPresidencia2021}, and it was not until 2022 that the first paper providing evidence on bias against the poor was published \cite{Curto2022}. Nonetheless, the authors are not aware of any studies that have unequivocally shown that aporophobia hinders poverty reduction.

This paper describes a computational approach to fill that gap, taking advantage of agent-based modelling. By means of simulation, the model presented here will allow testing numerous hypothetical legislation environments (i.e. the rules and laws in place) to predict their impact on poverty levels. In particular, we seek to answer the following question: how does the discrimination against the poor inherent in public policies affect outcomes concerning poverty levels and wealth distribution?

Our hypothesis is based on the social science literature, where there are indications that discrimination against the poor can become an obstacle to reduce poverty. The concept of the ``undeserving poor'' explains why governments have difficulty in passing laws to mitigate poverty when the poor are being criminalized for their situation \cite{Applebaum2001,Arneson1997,Nunn2009,Everatt2009}. In addition, on a personal level, it is more difficult to come out of poverty when you feel stigmatized by society \cite{Taylor2009}. 

The persistent criminalization of the poor denounced by NGOs can be explained by the shared meritocratic beliefs described by Sandel \cite{Sandel2020a} and the rhetoric of equal opportunity \cite{Fishkin2016}, which overestimates the responsibility of the poor for their situation \cite{Fraser2003}. However, the studies performed by the United Nations clearly state that the poor are overwhelmingly those born into poverty \cite{UnitedNations2018}. Chancel and Piketty report increasing numbers of global income inequality \cite{Chancel2021}, while intergenerational social mobility only corresponds to 7\% both in the US \cite{Chetty2014} and in Europe \cite{OECD2018}. In addition, in the US there is an overestimation of social mobility \cite{Alesina2018}, which exacerbates, even more, the blame put on the poor.

In the remainder of this paper, we describe our proposal to build an aporophobia agent-based model (AABM). ABMs allow studying the potential effects of innovative public policies in a society of interacting agents before they are implemented in a real-life society. As such, they are powerful tools to understand the efficacy and possible side effects of new regulations. In this paper, we first explain the agent decision-making model for the AABM based on previous needs-based modelling literature. Then we explain how we model agents' needs, actions and interactions with the physical and regulatory environment, which is central to the AABM. Aporophobia is a complex topic that can be analysed at a personal level in the relationship among agents and at a structural or institutional level. In our proposal, aporophobia is modelled at an institutional level through the poverty mitigation policies in place in a first case study based on the city of Barcelona. Finally, we describe the next steps to deliver empirical results, based on the use of the AABM, to evidence whether discrimination against the poor constitutes an obstacle to poverty reduction.

\section{An Agent-Based Model for Aporophobia}\label{sec:model}
\subsection{Overview}\label{subsec:model-overview}

\begin{figure}[th]
    \centering
    \includegraphics[width=0.95\linewidth]{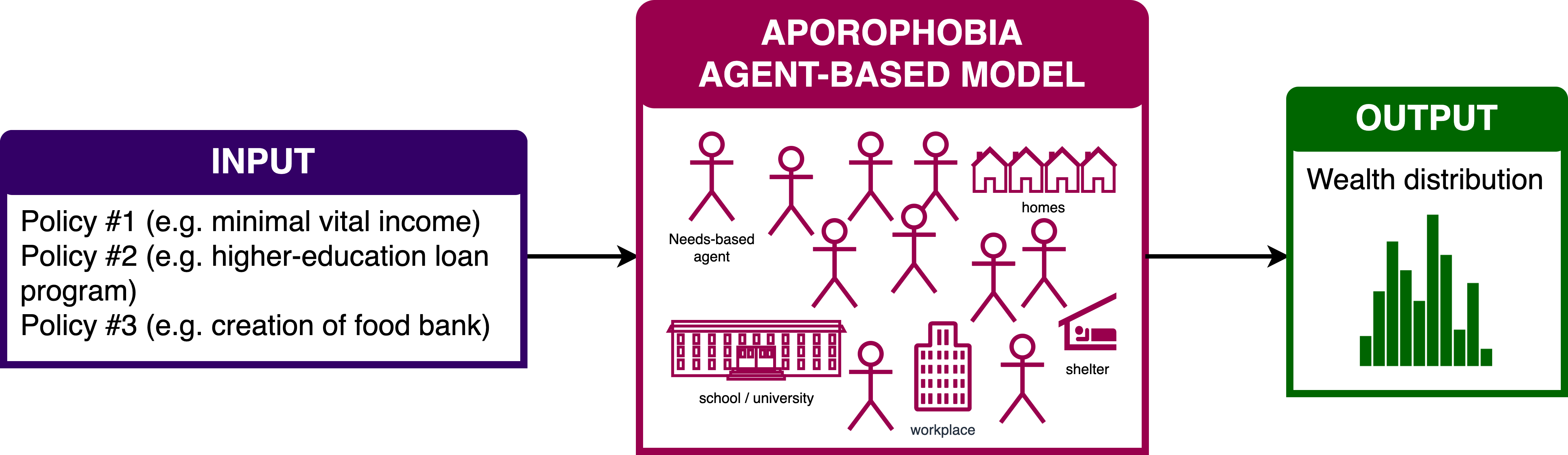}
    \caption{Overview of our pipeline.}
    \label{fig:overview}
\end{figure}

An overview of the pipeline of our work is presented in \Cref{fig:overview}. As an input, real-world policies that illustrate aporophobia at an institutional level are selected and included in the AAMB. We refer to the set of policies being applied as the \emph{regulatory environment}. The agent-based model is composed of a society of agents that interact within a shared environment that reflects features of everyday life, such as schools, homes, workplaces and other. The agents populating the model follow their strategies according to the needs-based model by Dignum et al., which we review in the following section. Poverty is a multidimensional phenomenon described as the deprivation of basic capabilities to live with dignity \cite{Sen2001}. Outputs resulting from the different simulations (modifying the AABM regulatory environment) will allow identifying the impact that specific policies (labelled according to their level of discrimination) have on poverty levels by observing the agents needs and actions as well as wealth distribution.

\subsection{The Agent Model}\label{subsec:agent-model}
We formulate our agent decision-making model following the needs-based model proposed by Dignum et al. This model was originally implemented to examine the impact and efficacy of non-medical interventions (e.g. mask mandates, lockdowns) during the Covid-19 pandemic \cite{Dignum2021,Dignum2020}. We adapt their model to the socio-economic problem (poverty) we are examining and adhere to the original spirit of the model formulation.

In the needs-based model, agents are characterized by their \emph{profile}, i.e. the set of relevant demographic and socio-economic characteristics. For example, in our model we consider variables such as age, gender, address and income. In addition, agents' internal states are described in terms of a set of \emph{needs} that are grouped into \emph{need categories}, according to Maslow's hierarchy of needs \cite{Maslow1943}. These needs and the importance that agents assign to them capture the agents' values and the motives that drive them to action. Needs have a certain satisfaction level that decays with time unless the agent takes action or finds itself in new states that refill a need. For example, an agent with a low level of satisfaction with the need ``belonging'' will not refill it unless it spends some leisure time with family and friends.

Mathematically, we denote the set of \emph{need categories} as $\mathcal{C}$, and the set of needs for category $c \in \mathcal{C}$ as $\mathcal{N}_c$. An \emph{importance function} $\mathsf{Imp}$ maps every need category to the weight that the agent assigns to it, $\mathsf{Imp}: \mathcal{C} \rightarrow [0,1]$. At time-step $t$, the need satisfaction level of need $n \in \mathcal{N}_c$ (for some $c \in \mathcal{C}$) is given by the \emph{need satisfaction level function} $\mathsf{NSL}_t : \{\bigcup_{c \in \mathcal{C}} \mathcal{N}_c\} \rightarrow [0,1]$, which maps every need (across all categories) to its current degree of fulfillment. Conversely, the \emph{urgency} of need $n$ at time step $t$ is defined as $\mathsf{Urg}_t(n) = 1 - \mathsf{NSL}_t(n)$.

Agents perform actions with the intention of refilling their most urgent needs, i.e. those which are the most depleted. We denote the set of actions available to an agent by $\mathcal{A}$. These actions are (possibly) determined by the current state and/or profile of the agent (e.g. only an employed agent can currently take action ``go to workplace'' or ``work''). To decide on which action to take next, agents consider the satisfaction they expect to draw from the execution of that action. The \emph{expected satisfaction function} $\mathsf{Sat}: \{\bigcup_{c \in \mathcal{C}} \mathcal{N}_c\} \times \mathcal{A} \rightarrow [0,1]$ captures these estimations. Hence, $\mathsf{Sat}(n, a)$ indicates the level of satisfaction the agent expects to get for need $n$ after executing action $a$.

At every time step $t$, the agent deliberates about what is the best action $a_t$ to perform at the current time step, considering its available actions and its needs state. For every action, the agent assigns a score based on the level of satisfaction it is going to draw from executing it, across all of its needs weighted by the current urgency and the importance of their category. Once all available actions have been scored, the agent follows a greedy policy and selects the action with the largest score to perform. Formally, we propose the deliberation function to follow the expression:
\begin{equation}\label{eq:deliberation-function}
    a_t = \argmax_{a \in \mathcal{A}} \left[\sum\limits_{c \in \mathcal{C}} \left( \sum\limits_{n \in \mathcal{N}_c} \mathsf{Sat}(a,n) \cdot \mathsf{Urg}(n) \right) \cdot \mathsf{Imp}(c)  \right]
\end{equation}
Hence, agent strategies are determined by the importance they assign to their needs and the satisfaction they expect to draw from the execution of their action. Since poverty is a relative concept, we will measure the level of poverty by the changes in the agents fulfilment of needs such as housing, education and food. 

Although agents perform actions with the intention of improving their needs state, it is not a guarantee that at the next time step they will be satisfied according to the expected degree. Hence, after \emph{all} agents have executed their selected actions, $\mathsf{NSL}_{t+1}(n)$ is computed for all the needs of every agent according to the new state of the system. For example, a homeless person might choose to go to a community centre to look for shelter. However, if there are no beds available in the community centre, the need for shelter will not be fulfilled.

\begin{figure}[t]
    \centering
    \includegraphics[width=0.9\linewidth]{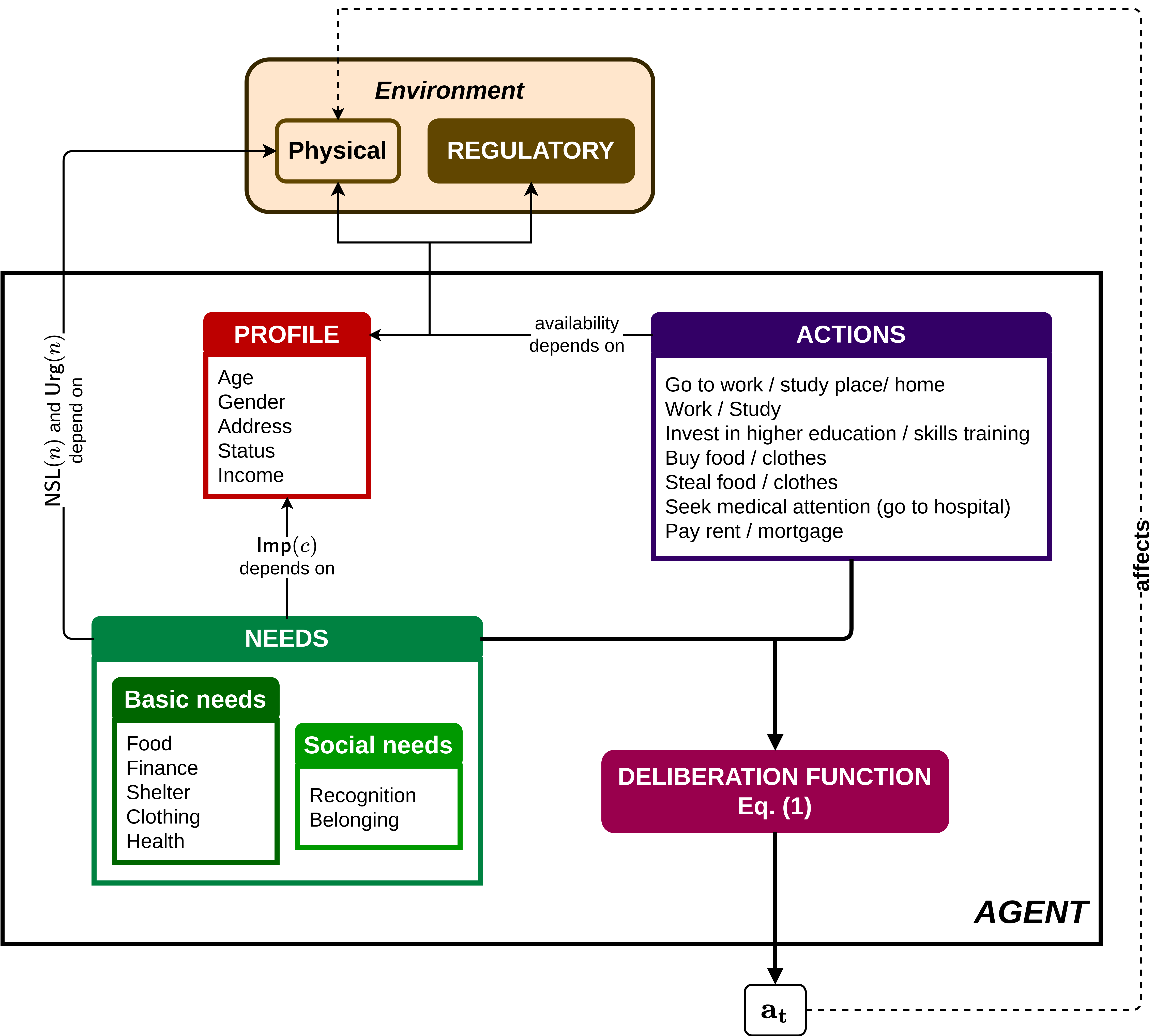}
    \caption{Architecture of the AABM, aporophobia agent based model, based on the needs model of Dignum et al.}
    \label{fig:our-model}
\end{figure}

The proposal for the needs-based agent model of our AABM appears in \Cref{fig:our-model}. In it, we outline what are the specific profile features, needs and actions that we consider for our AABM. First, we consider profile features related to basic demographic information (age and gender) and socio-economic factors representing the housing address (for a homeless agent, the address would equal to \emph{null}), income and status (student, employed, unemployed or retired, independently of income).

In line with the multidimensionality of poverty, we also model the needs of agents using two categories: basic and social needs. Basic needs include access to food, financial security, shelter, clothing, health state and education. Although people who have a secure position have all of these needs relatively well satisfied, in situations of poverty (the phenomenon we are examining) this may not be the case. Social needs include recognition by the rest of society (which is expressed as having a valuable job, caring responsibilities or increasing one's education or skills level) and belonging (which is expressed as spending time with loved ones). We intend to model social needs in order to examine the impact of poverty on the psychological well-being of those trapped by it, and how poverty-mitigating measures can, albeit indirectly, have an effect on it.

Third, we propose to model a basic set of actions related to an agent's daily activities, like going to the workplace or place of study, returning home, paying for food, clothing and rent or mortgage. We also include an \emph{investment} activity in the form of payment for higher education and \emph{criminal} activities such as stealing goods. Finally, we also consider the case where agents need to seek medical attention due to the low fulfillment of their health need.

The relationships between the components of our model are also outlined in \Cref{fig:our-model}. Namely, action availability depends on the agent profile and the current state of the environment, both physically (e.g. if an agent works in a factory, it can only perform action ``work'' once it is there), and on the regulations that are in place (e.g. access to higher education for low-income agents depends on the availability of grants or loans). Concerning needs, we consider that the importance of need categories $\mathsf{Imp}(c)$ depends mostly on the agent profile, while their satisfaction level (and therefore, the urgency to refill them) depends on the current state. The needs state and the available actions are considered by the agent to deliberate about which action to perform next, which in turn affects the state of the environment.

\subsection{The Regulatory Environment}\label{subsec:regulation}
\Cref{fig:our-model} also outlines how the different components of our agent model interact with the \emph{environment} where agents are situated. We consider this environment to be composed of two elements. The first element is the \emph{physical} environment. This element encompasses the classical environmental features of an \acrshort{abm}, i.e. the (virtual) space where agents inhabit, interact, and take actions.

The second element is the \emph{regulatory} environment where agents are situated. This is composed of the policies, laws and regulations that are implemented within the system. We make several assumptions concerning these policies. First, we assume that they are perfectly implemented. Thus, we do not introduce any agents (such as social workers) that are responsible for the implementation of policies. Second, we assume that all agents in the model are perfectly aware of the rules in place. In particular, they know about potential new actions that they may take (such as applying for housing or citizen income).

Third, we assume that the agent's actions are unable to affect the set of policies in place. This means that, for this version of the ABM, we do not model any voting or legislative process in which agents could change the rules that affect them. We make this assumption at this point because we wish to observe the \emph{effect} that certain policies have on poverty, and not model the policy-making process itself. The assumptions we have presented so far can be summarized as abstracting away any government bodies from the \acrshort{abm}, meaning there are no (potentially imperfect) agents in charge of implementing policies endogenously, and the processes foreseen by any regulation proceed exactly as stated.

The regulatory environment has a very prominent role in our \acrshort{abm} because it is through the policies in place that \emph{aporophobia at the institutional level} is modeled. We do not model aporophobia as an individual agent construct, since there is not an established body of psychological research on bias against the poor to back such modeling. However, we assume, according to the inputs received by NGOs specialized on poverty, that ``institutional aporophobia'' is reflected in the laws and rules that are implemented in society. This is the reason why the regulatory environment is a central piece of our model: we use the rules and regulations that are contained in it as \emph{proxies} for aporophobia. In other words, we translate value interpretations into value representations as norms \cite{Noriega2022}. Since poverty is modelled as a diversity of agents' needs, we will be able to observe the impact that changes in aporophobic regulations have on poverty levels by identifying changes in these needs.  

In order to systematically collect the policies that we wish to implement, we intend to use the \acrfull{ig}, originally presented by Crawford and Ostrom \cite{Crawford1995} and recently expanded into its 2.0 version by Frantz and Siddiki \cite{Frantz2020,Frantz2022}. In its coarsest version, the \acrshort{ig} parses institutional statements into the \textbf{ADICO} syntax: the \textbf{A}ttribute field indicates to whom the convention, norm or rule applies; the \textbf{D}eontic indicates whether it is a prohibition, permission or obligation; the a\textbf{I}m field indicates what action or outcome the deontic applies to; the \textbf{C}ondition field indicates under what circumstances the statement applies; and the \textbf{O}r-Else field indicates any consequences for detected violations.

In \Cref{tab:example-rules-parsing}, a sample of three regulations in force in the city of Barcelona illustrate how policies are going to be described according to the \acrshort{ig}. In the next steps developing the AABM, a richer representation of policies of different level of governance will be included. In addition to the parsed fields anticipated by the \acrshort{ig}, rules will be annotated by the larger policy body they belong to (i.e. the law or regulation) and the level of government (i.e. supra-national, national, regional or local). Furthermore, a final annotation will be introduced with the degree of \emph{discriminatory} or \emph{distributive} character of the rule. For this point, we have secured the help of several foundations and NGOs working on poverty alleviation, who have solid in-the-field experience navigating the laws and regulations that we intend to model. Their input will be the main source for annotating rules by their discriminatory or distributive degree. The rules annotations (indicating origin and aporophobic degree) will allow examining the interplay between policies along these two dimensions.

\begin{table}[b]
    \centering
    \caption{Sample regulations in force in the city of Barcelona, according to the various jurisdictions the city is under.}
    \label{tab:example-rules-parsing}
    \begin{tabular}{|c|l|}
        \hline \textbf{Reference} & \cite{GobiernodeEspana2015} \\
        \hline \textbf{Jurisdiction} & National \\
        \hline \textbf{A}ttribute & anyone \\
        \hline \textbf{D}eontic & must \\
        \hline A\textbf{I}m & pay fine of 100€ to 600€ \\
        \hline \textbf{C}ondition & if they misuse public furniture (e.g. sleep on a bench) \\
        \hhline{|=|=|}
        \hline \textbf{Reference} & \cite{ConsorciDHabitatgedeBarcelona2016} \\
        \hline \textbf{Jurisdiction} & Regional \\
        \hline \textbf{A}ttribute & anyone \\
        \hline \textbf{D}eontic & can \\
        \hline A\textbf{I}m & enter the social emergency program \\
        \hline \textbf{C}ondition & if they have lost their home \\
        \hhline{|=|=|}
        \hline \textbf{Reference} & \cite{JefaturadelEstado2021} \\
        \hline \textbf{Jurisdiction} & National \\
        \hline \textbf{A}ttribute & anyone with home address and residency in Spain, and a bank account \\
        \hline \textbf{D}eontic & can \\
        \hline A\textbf{I}m & apply for minimal vital income \\
        \hline \textbf{C}ondition & -- \\
        \hline
    \end{tabular}
\end{table}

\subsection{Use Case}\label{subsec:uses}
Poverty is a phenomenon affected not only by socioeconomic factors but also by educational, cultural and institutional contexts \cite{Sen2001}. Therefore, a case study approach, starting from a specific city, allows considering the specificity of the context while working towards providing the first empirical evidence (at a small scale) of whether aporophobia constitutes an obstacle to mitigate poverty. 

As a first case study, we have selected poverty reduction regulations enforced in the city of Barcelona, due to the contribution of local NGOs to identifying the specific laws (at a local, regional and national level) that are included in the computational model. The sample policies presented in \Cref{tab:example-rules-parsing} have been parsed into their components according to the coarsest level of the \acrshort{ig}. The \textbf{O}r-Else field (see \Cref{subsec:regulation}) has been omitted because none of the examples in \Cref{tab:example-rules-parsing} has a provision for sanctions.

Once the first results are obtained for the first case study, fieldwork will be conducted in collaboration with local NGOs and the municipality of Barcelona. In addition, AABM will be enriched with the laws and regulations for the cities of Johannesburg and New York. The rationale for the selection of the city of Johannesburg obeys the interest to work hand in hand with NGOs and computer scientists in South Africa. Current global inequality in AI development contributes to the concentration of profits in the Global North and the danger of ignoring the context in which AI is applied. In practice, Chan et al. document that out of the top 100 universities and companies by publication index, none of them are from Africa or Latin America \cite{AlanChan2021}. Finally, the third case study will be based on New York since it is the first city to report to the UN regarding the SDGs \cite{NYCMayorsOfficeInternationalAffairs} in a context where a persistent criminalization of poverty \cite{UnitedNations2018} and an overestimation of social mobility \cite{Alesina2018} have been documented.

\section{Conclusions and Next Steps}\label{sec:conclusions}
We have presented the architecture for an aporophobia agent-based model (AAMB). We have worked with local NGOs to identify enforced laws that act as proxies for institutional aporophobia in the case study for the city of Barcelona. Then we have presented a sample of how these policies will be described in the regulatory environment of the AAMB by using the \acrshort{ig}.

Next steps will focus on obtaining experimental results from the AABM that provide evidence of whether aporophobia has an influence on actual poverty levels. For that purpose, the model will be enriched with a wider set of policies that in-the-field experts consider discriminatory against the poor in the case study. The different levels of governance and their interaction will be described. Enforced policies that have an impact on economic indicators such as income inequality, GDP growth and taxation will also be considered, since preliminary studies suggest that these indicators could be correlated to aporophobia. 

The main objective of this research is to mitigate poverty and discrimination in the real world. Therefore, we are working with local NGOs and it is foreseen to use the results of this research to evaluate changes in the regulatory environment in collaboration with the municipality of Barcelona. For that purpose, real data collected from fieldwork will be incorporated in the AAMB in order to track the impact of the regulatory changes on the homeless people. At a second stage, the AABM will incorporate the sets of policies applicable to the cities of Johannesburg and New York. 

The empirical results obtained from the AABM have the potential to inform a new generation of poverty reduction policies that act not only on the redistribution of wealth, but also on mitigating discrimination against the poor. This exploratory path to fight against poverty puts the focus not only on the poor, but on society at large.


\begin{acknowledgments}
We would like to thank Beatriz Fernández, head of the legal team in Fundació Arrels (\url{www.arrelsfundacio.org}) for her contributions to identifying the legal framework related to the topic of aporophobia in the case study for the city of Barcelona.

This work has been supported by the EU funded VALAWAI (Proj. No. 101070930) and WeNet (Proj. No. 823783) projects, and the Spanish funded VAE (Proj. No. TED2021-131295B-C31) and Rhymas (Proj. No. PID2020-113594RB-100) projects.
\end{acknowledgments}

\bibliography{references}

\end{document}